\documentclass[10pt]{IEEEtran}
\IEEEoverridecommandlockouts
\usepackage{cite}
\usepackage{amsmath,amssymb,amsfonts}
\usepackage{graphicx}
\usepackage{textcomp}
\usepackage{setspace}
\usepackage{booktabs}
\usepackage{multirow}
\usepackage{hyperref}
\usepackage{url}
\usepackage{nicefrac}
\usepackage{microtype}
\usepackage{xcolor}
\usepackage{float}
\usepackage{orcidlink}
\usepackage{comment}
\usepackage{braket}

\def\BibTeX{{\rm B\kern-.05em{\sc i\kern-.025em b}\kern-.08em
    T\kern-.1667em\lower.7ex\hbox{E}\kern-.125emX}}

\usepackage{enumitem}
\setlist[itemize]{leftmargin=*}%
\setlist[enumerate]{leftmargin=*}%

\usepackage[compact]{titlesec}
\usepackage{stfloats}

\titlespacing\section{0pt}{0.2\baselineskip}{0.12\baselineskip}
\titlespacing\subsection{0pt}{0.15\baselineskip}{0.08\baselineskip}
\titlespacing\subsubsection{0pt}{0.1\baselineskip}{0.08\baselineskip}

\makeatletter
\def\bstctlcite{\@ifnextchar[{\@bstctlcite}{\@bstctlcite[\@IEEEstringA]}}
\def\@bstctlcite[#1]#2{\gdef\@IEEEcurrentbstctl{#2}}
\IEEEtriggercmd{\reset@font\normalfont\refsize} 
\IEEEtriggeratref{1} 
\makeatother

\usepackage{fancyhdr,lipsum}
\fancypagestyle{firstpage}{
  \fancyhf{}
  \fancyhead[C]{To appear at the IEEE International Conference on Quantum Artificial Intelligence (QAI), Nottingham, UK, December 2026.}
  \fancyfoot[C]{\thepage}
}

\begin{document}
\pagestyle{empty}

\title{QLIF-CAST: Quantum Leaky-Integrate-and-Fire for Time-Series Weather Forecasting
}


\author{\IEEEauthorblockN{Alberto Marchisio\IEEEauthorrefmark{4}\IEEEauthorrefmark{2}\IEEEauthorrefmark{3}$^*$\thanks{$^*$This work has been carried out while at NYUAD.}, Aayan Ebrahim\IEEEauthorrefmark{4}\IEEEauthorrefmark{2}, Nouhaila Innan\IEEEauthorrefmark{4}\IEEEauthorrefmark{2}, Muhammad Kashif\IEEEauthorrefmark{4}\IEEEauthorrefmark{2}, Muhammad Shafique\IEEEauthorrefmark{4}\IEEEauthorrefmark{2}}

\IEEEauthorblockA{\IEEEauthorrefmark{4} \normalsize eBrain Lab, Division of Engineering, New York University Abu Dhabi, PO Box 129188, Abu Dhabi, UAE\\}
\IEEEauthorblockA{\IEEEauthorrefmark{2} \normalsize Center for Quantum and Topological Systems, NYUAD Research Institute, New York University Abu Dhabi, UAE\\}
\IEEEauthorblockA{\IEEEauthorrefmark{3}
\normalsize SDU Microelectronics, Institute of Mechanical and Electrical Engineering, University of Southern Denmark, Odense, Denmark\\
Emails: chisio@sdu.dk, \{ae2417, nouhaila.innan, muhammadkashif, muhammad.shafique\}@nyu.edu}
\vspace{-25pt}
}

\thispagestyle{empty}
\maketitle
\thispagestyle{empty}
\pagenumbering{gobble}
\thispagestyle{firstpage}

\begin{spacing}{0.92}

\begin{abstract}

Accurate and efficient time-series forecasting remains a challenging problem for both classical and quantum neural architectures, particularly in multivariate environmental settings. This work adapts the Quantum Leaky Integrate-and-Fire (QLIF) spiking neural network for time-series regression tasks, specifically short-term multivariate weather forecasting. We extend QLIF beyond classification and demonstrate its applicability to continuous-valued prediction problems.

The QLIF-CAST model encodes neuron excitation states as single-qubit quantum superpositions, driven by $R_x$ rotation gates and T1 relaxation decay, and is embedded within a hybrid quantum-classical recurrent architecture. We conduct two distinct evaluations. First, a controlled comparison against a parameter-matched classical LIF baseline on a multivariate weather dataset shows that QLIF-CAST achieves 15.4\% lower MSE and 4.4\% lower MAE, demonstrating that quantum neuronal dynamics reduce prediction error over classical equivalents. Second, a cross-domain comparative analysis with state-of-the-art quantum LSTM (QLSTM) and quantum neural network (QNN) models on air quality and wind speed benchmarks reveals that QLIF-CAST converges in up to 94\% less training time, occupying a distinct position in the speed-error trade-off space. Hardware verification on IBM Marrakesh (156-qubit QPU) confirms reliable circuit execution with only 1.2\% average deviation from simulation.

\end{abstract}

\begin{IEEEkeywords}
Quantum Machine Learning, Spiking Neural Networks, Time-Series Forecasting, Weather Prediction, Quantum-Classical Hybrid
\end{IEEEkeywords}
\vspace{-10pt}
\end{spacing}
\begin{spacing}{0.99}
\section{Introduction}

Short-term multivariate weather forecasting requires models capable of capturing complex temporal dependencies across correlated physical variables~\cite{xingjian2015convolutional,hewage2020temporal}. Traditional recurrent architectures such as LSTM often trade computational cost for predictive quality~\cite{hochreiter1997lstm,greff2017lstm}, while classical spiking neural networks offer event-driven efficiency but lack expressive state representations~\cite{maass1997networks,pfeiffer2018deep}.
Quantum spiking models combine the discrete, event-driven nature of spikes with quantum superposition and interference, potentially offering richer state representations than classical neurons within the same parameter budget~\cite{kristensen2021artificial,konar2021qspike}. However, while the QLIF model~\cite{innan2024quantum, brand2024QLIF} has demonstrated strong results on image classification (MNIST, 98.1\%), its application to time-series regression remains unexplored.
In this paper we address this gap through two complementary evaluations:

\begin{enumerate}
    \item \textbf{Controlled quantum vs.\ classical comparison:} QLIF-CAST is compared against a parameter-matched classical LIF baseline, identical in all respects except the neuronal update rule. This isolates the effect of quantum dynamics on prediction error for multivariate weather forecasting.
    \item \textbf{Comparative analysis against published quantum models:} QLIF-CAST is evaluated on benchmark tasks used by QLSTM~\cite{bangkok_qlstm} and LSTM-QNN~\cite{qnn_wind}, characterizing the speed-error trade-off between simple single-qubit circuits (in QLIF-CAST) and deeper variational quantum circuits (in QLSTM and LSTM-QNN).
\end{enumerate}

\textbf{Contributions:} This work makes the following contributions:
\begin{enumerate}
    \item \textbf{Novel application domain for QLIF:} We extend the QLIF spiking neuron, originally demonstrated on image classification, to continuous-valued multivariate time-series regression, establishing it as a viable architecture for temporal forecasting tasks.
    \item \textbf{Systematic speed-error characterization:} We provide the first systematic comparison of QLIF-CAST's depth-2, classically-driven single-qubit circuits (with no trainable quantum parameters) against deeper variational quantum architectures (QLSTM, LSTM-QNN) across multiple real-world forecasting benchmarks, quantifying the trade-off between training efficiency and prediction accuracy.
    \item \textbf{Cross-domain benchmark evaluation:} QLIF-CAST is evaluated across three distinct environmental forecasting domains (meteorological, air-quality, and wind-energy) without modifying the core quantum circuit, demonstrating architectural generality.
    \item \textbf{Practical vectorized implementation:} We introduce a batched, vectorized circuit execution strategy in which all rotation angles across all neurons and timesteps in a batch are executed in a single parallel circuit call, yielding a $500{\times}$ speedup over sequential execution and making QLIF-CAST training on real-scale datasets feasible under classical simulation.
    \item \textbf{Near-term QPU hardware validation:} We provide empirical validation of the QLIF-CAST circuit on a real 156-qubit QPU (IBM Marrakesh), contributing a reproducible hardware benchmark for shallow, classically-driven quantum circuits in the QML literature.
\end{enumerate}

\section{Background and Related Work}

\subsection{Spiking Neural Networks for Time-Series}

Spiking neural networks (SNNs) process information through discrete spike events rather than continuous activations, offering a computational model closer to event-driven neural dynamics than conventional artificial neural networks~\cite{gerstner2002spiking}. Among the most widely used spiking neuron models, the Leaky Integrate-and-Fire (LIF) neuron combines input integration, membrane leakage, and threshold-triggered spike emission in a simple dynamical system~\cite{izhikevich2004model, Wang2014LIF}. Because of this balance between simplicity and temporal modeling capability, LIF-based architectures have become a common starting point for sequence processing in spiking systems. 

Prior work has shown that SNNs can be effective for temporal tasks such as speech recognition and event-driven sensor processing~\cite{dong2021deep, kim2020spiking}. However, most existing studies emphasize classification or detection, while comparatively fewer investigate continuous-valued forecasting. In particular, the use of SNNs for multivariate regression remains limited, leaving open the question of whether spike-based dynamics can support accurate prediction in practical time-series settings.

\subsection{Quantum Neural Networks}

Quantum neural networks aim to exploit quantum-mechanical effects such as superposition and entanglement to construct models with richer internal representations than their classical counterparts~\cite{cerezo2021variational}. A common approach is to use variational quantum circuits (VQCs), in which parameterized quantum gates are optimized with classical gradient-based routines~\cite{cong2019quantum}. Such hybrid quantum-classical models have shown promise in the noisy intermediate-scale quantum (NISQ) setting, where near-term quantum devices are available but remain constrained by limited depth and noise sensitivity~\cite{bharti2022noisy}. 

However, VQC-based models face practical limitations. As circuit depth grows, gate noise accumulates and the cost of training increases, particularly when gradients are estimated through parameter-shift rules. In recurrent settings, where quantum subcircuits may be invoked repeatedly across time steps, this overhead can become a substantial bottleneck. These considerations motivate interest in quantum architectures that retain useful quantum dynamics while avoiding the training cost of deeply parameterized circuits.

\subsection{Classical Leaky Integrate-and-Fire}
\label{subsec:classical_LIF}

\begin{figure}[t!]
\centerline{\includegraphics[width=\columnwidth]{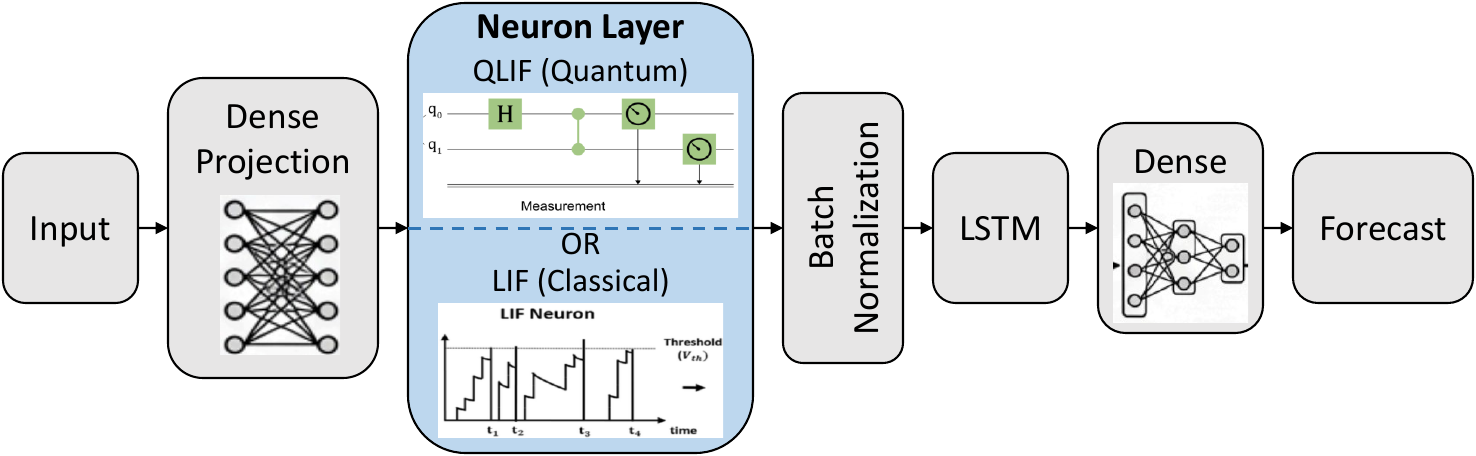}}
\caption{Architecture using Quantum LIF or Classical LIF neuron model.
\textbf{Classical LIF}: The membrane potential $U$ integrates synaptic input $I_{\text{in}}$ and decays exponentially between spikes ($U_{\text{new}} = \beta U_{\text{prev}} + (1{-}\beta)I_{\text{in}}$, $\beta = e^{-1/\tau}$). A spike is emitted when $U$ exceeds the threshold, after which $U$ resets to zero. \textbf{Quantum LIF (QLIF)}: The qubit is initialized in $|0\rangle$, rotated by $R_x(\phi)$ to encode the current excitation state, then rotated by $R_x(\theta_{\text{input}})$ to integrate the input signal. Measuring the qubit yields the updated excitation probability $\alpha_{\text{new}} = \sin^2(\frac{\phi + \theta_{\text{input}}}{2})$.}
\label{fig:qlif_lif_architecture}
\end{figure}

As shown in Fig.~\ref{fig:qlif_lif_architecture}, a model can employ either the Classical Leaky Integrate-and-Fire (LIF) neuron model or the Quantum LIF (QLIF). The Classical LIF neuron models membrane potential dynamics via an exponential decay rule and input accumulation~\cite{Wang2014LIF}. The membrane potential $U$ integrates incoming current and decays toward rest between spikes. A spike is emitted when $U$ exceeds a fixed threshold, after which the potential resets. This model will serve as the baseline for Phase~1 of our study.

\subsection{QLIF Neuron Model}

The Quantum Leaky Integrate-and-Fire (QLIF) model~\cite{brand2024QLIF} adapts the logic of spiking neurons to a single-qubit setting by representing neuronal excitation as the probability of measuring the qubit in the state $|1\rangle$. At each time step, the excitation state is updated through $R_x$ rotations that encode the current state and the incoming signal, yielding a compact quantum version of state accumulation and decay. Unlike standard variational quantum models, the circuit does not introduce separately trainable quantum gate parameters. Instead, the rotation angles are computed directly from classical network outputs. 

This design choice is important for two reasons. First, it preserves a simple quantum circuit structure that is easier to execute and analyze than deeper variational alternatives. Second, it avoids the additional optimization burden associated with trainable quantum parameters, making QLIF closer in training complexity to a classical recurrent component than to a fully variational quantum layer. Prior work has demonstrated the effectiveness of QLIF in image classification, which motivates examining whether the same neuron model can support continuous-valued temporal prediction.

\subsection{QLSTM and LSTM-QNN}
\label{sec:comparison_models}

In addition to the controlled classical baseline, Phase~2 compares QLIF-CAST against two state-of-the-art quantum forecasting models (QLSTM and LSTM-QNN) that use more expressive but more expensive variational quantum components. These models provide useful reference points for understanding where QLIF-CAST lies in the trade-off between predictive accuracy, architectural complexity, and training cost.

\begin{figure}[t!]
\centerline{\includegraphics[width=\columnwidth]{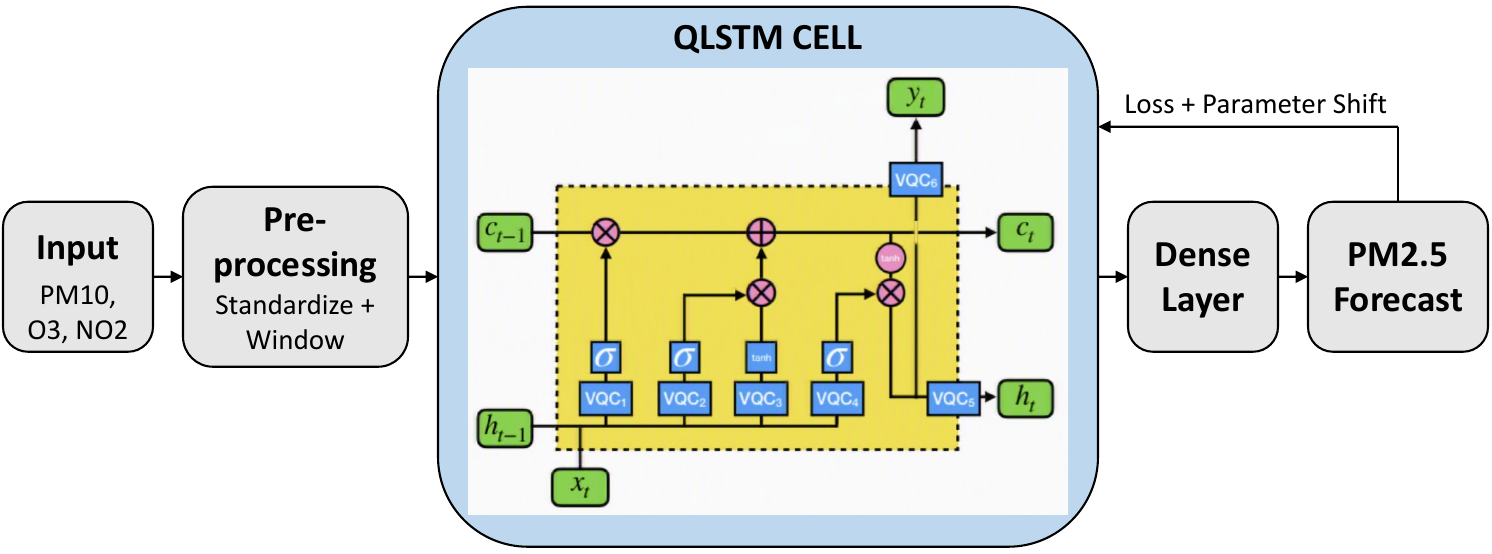}}
\caption{QLSTM architecture. Each of the four classical LSTM gates (forget, input, cell, output) is replaced by a multi-qubit variational quantum circuit (VQC), with gate angles trained via parameter-shift gradient rules.}
\label{fig:qlstm_arch}
\end{figure}

\textbf{QLSTM}~\cite{bangkok_qlstm} replaces the gates of a classical LSTM with parameterized multi-qubit variational quantum circuits, as illustrated in Fig.~\ref{fig:qlstm_arch}. In this design, the forget, input, cell, and output transformations are each implemented through trainable quantum subcircuits optimized by parameter-shift gradients. Reported results show an MAE of 11.24~$\mu$g/m$^3$ on Bangkok PM2.5 forecasting after 100 training epochs. While this architecture can offer strong predictive performance, its repeated use of multi-qubit variational blocks makes training substantially more expensive than that of a standard classical LSTM.

\begin{figure}[t!]
\centerline{\includegraphics[width=\columnwidth]{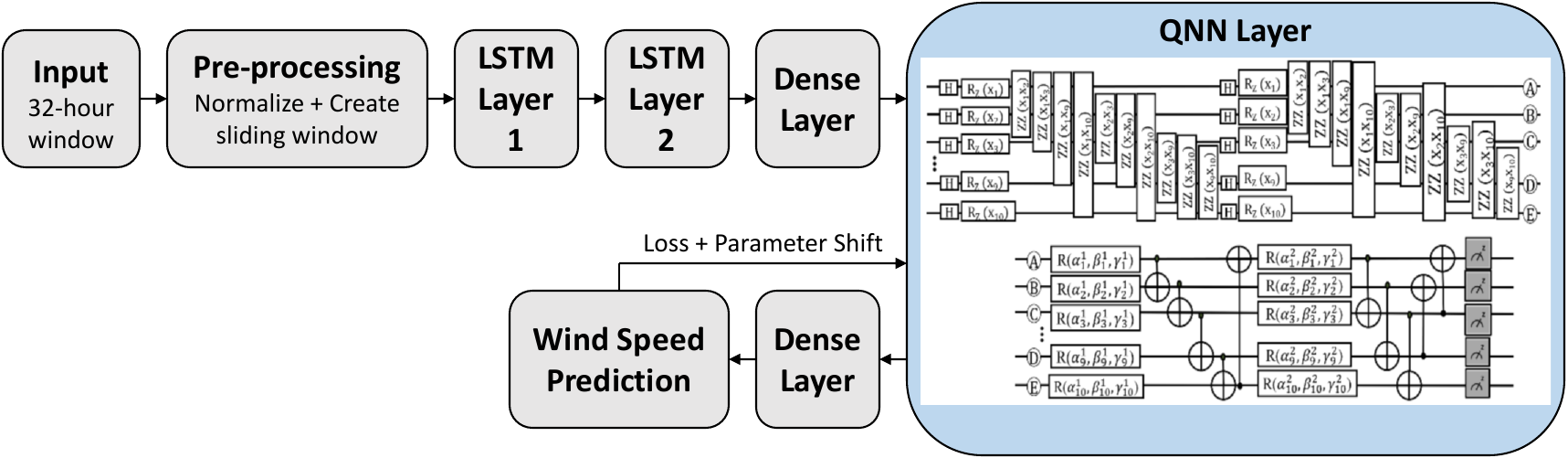}}
\caption{LSTM-QNN architecture. A classical LSTM encoder performs temporal feature extraction, whose output is passed to a variational quantum neural network (QNN) before the regression head.}
\label{fig:lstmqnn_arch}
\end{figure}

\textbf{LSTM-QNN}~\cite{qnn_wind} follows a different hybrid strategy. As shown in Fig.~\ref{fig:lstmqnn_arch}, a classical LSTM first extracts temporal features, after which the resulting representation is processed by a variational quantum neural network before the regression output is produced. This model achieves an RMSE of 3.92~km/h on hourly wind speed forecasting, with a reported training time of approximately 65 minutes. As in QLSTM, the variational quantum component increases computational cost through additional trainable circuit parameters and repeated quantum evaluations during optimization.

Taken together, these prior models highlight two contrasting directions in quantum sequence learning. QLSTM and LSTM-QNN pursue higher expressiveness through deeper variational quantum circuits, whereas QLIF-CAST emphasizes shallow, classically driven quantum state updates with no trainable quantum parameters. \textit{This distinction is central to our study, because it frames QLIF-CAST not only as an alternative quantum forecasting model, but also as a computationally lighter design within the broader quantum time-series literature.}

\section{QLIF-CAST Methodology}
\begin{figure*}[!t]
\centerline{\includegraphics[width=.95\linewidth]{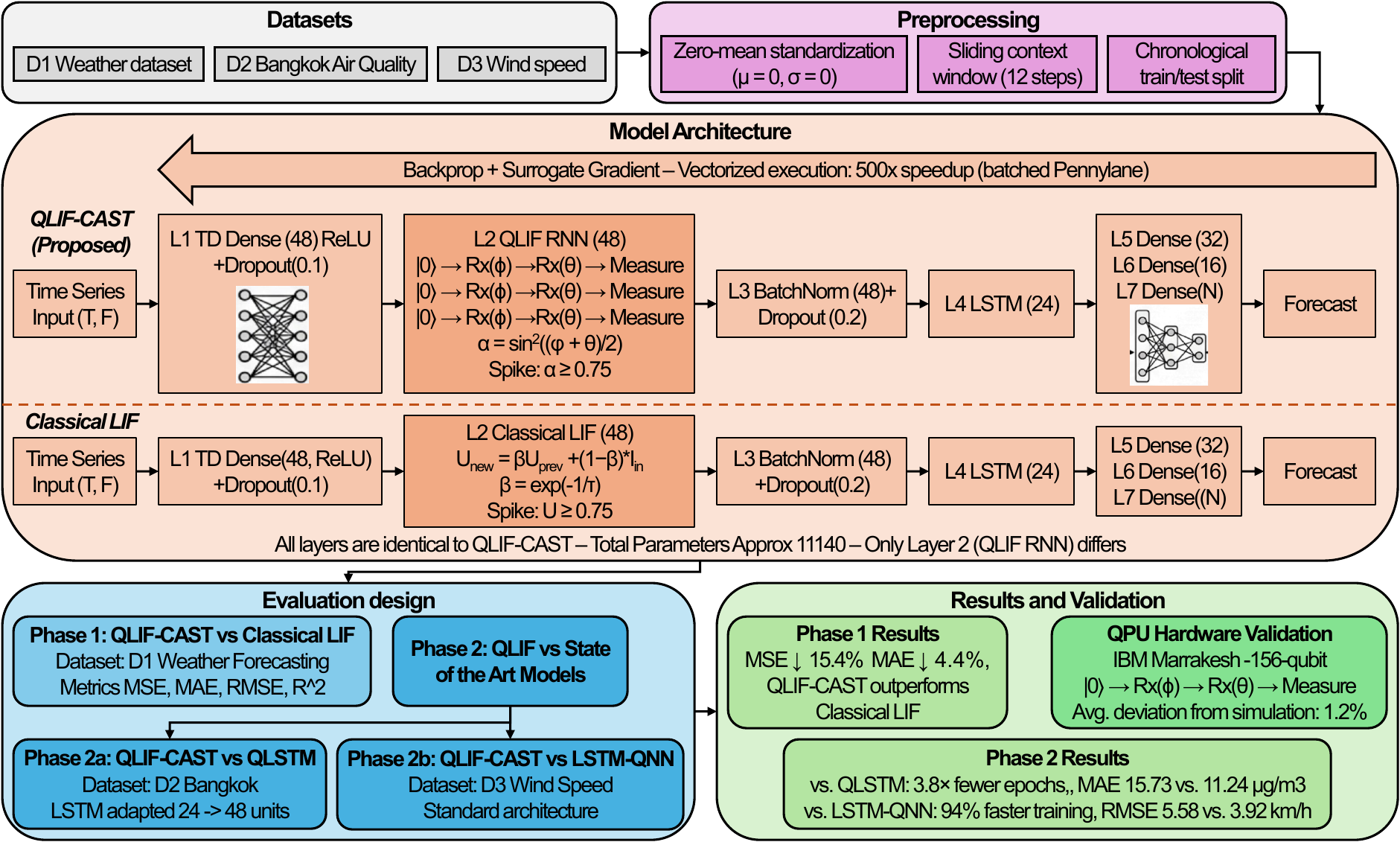}}
\caption{Overview of the QLIF-CAST methodology. Three datasets (D1-D3) used across the two evaluation phases, with shared preprocessing steps. QLIF-CAST and Classical LIF architectures; the two pipelines are structurally identical, while only Layer~2 (the neuronal update rule) differs. Two-phase evaluation design and key results, with QPU hardware validation on IBM Marrakesh.}
\label{fig:methodology}
\end{figure*}

This section describes the datasets, the QLIF neuron model, the hybrid network architecture, the matched classical baseline, and the experimental design. Fig.~\ref{fig:methodology} provides a high-level overview of the complete pipeline, covering data collection and preprocessing, both model architectures, the two-phase evaluation design, and QPU hardware validation.

\subsection{Datasets and Preprocessing}

We employ three distinct datasets, denoted as D1, D2, and D3, each tailored to a specific evaluation objective. Dataset D1~\cite{weather_kaggle} is utilized to assess the performance advantage of the proposed QLIF-CAST model over its classical counterpart (Classical LIF). Dataset D2~\cite{waqi} is used to benchmark QLIF-CAST against the state-of-the-art QLSTM. The rationale behind using a different dataset in this case is to ensure a fair comparison by aligning with the dataset used in the reference QLSTM study. Dataset D3~\cite{qnn_wind} is utilized to evaluate the performance of QLIF-CAST against the state-of-the-art LSTM-QNN model, with the dataset chosen to ensure consistency with the referenced study.
For all datasets, input features are standardized to have zero mean and unit variance to ensure stable and consistent training. Furthermore, a sliding context window mechanism is applied to transform the time-series data into supervised learning format, generating input-target pairs for model training and evaluation.
Below, we provide details of each dataset along with dataset specific preprocessing.

\subsubsection{Dataset D1 - Weather Forecasting}

The Weather History dataset~\cite{weather_kaggle} contains $96,453$ hourly observations spanning April $2006$ to September $2016$. Four continuous variables serve as both inputs and regression targets: Temperature~(°C), Relative Humidity~(fraction), Wind Speed~(km/h), and Atmospheric Pressure~(mb). A sliding window of 12~hours provides context, with the next-hour vector of all four variables as the regression target. The dataset is split chronologically into $10,000$ training and $2,000$ test samples.

\subsubsection{Dataset D2 - Bangkok Air Quality Dataset}

The WAQI dataset~\cite{waqi} provides daily air quality measurements at Bangkok monitoring stations from July $2016$ to January $2026$ ($\sim 3,290$~records). SO\textsubscript{2} and CO columns are excluded due to substantial missing data. To ensure a fair and direct comparison with the reference QLSTM study~\cite{bangkok_qlstm}, we adopt the same feature set and prediction target: three input features, PM10 ($\mu$g/m$^3$), O\textsubscript{3} (AQI units), and NO\textsubscript{2} (AQI units), are used to predict next-day PM2.5 concentration (the primary air quality target in the reference study, and the most health-critical pollutant for Bangkok). Short gaps ($\leq$3 days) are forward-filled (i.e., the last known value is carried forward to fill the gap); longer gaps are linearly interpolated; rows with remaining missing targets are dropped. A $12$-day context window is used, with an 80\%/20\% chronological train-test split.

\subsubsection{Dataset D3 - Wind Speed Dataset}

Hourly wind speed data at $100$m altitude is sourced from the Open-Meteo historical weather API~\cite{qnn_wind} containing $35,064$ observations from January $2020$ to December $2023$ at latitude $27.87$°N, longitude $40.14$°E (elevation $\approx$$1004$m). This altitude is directly relevant to wind turbine power generation. The series contains no missing entries. A $12$-hour context window targets wind speed at hour $t{+}1$, with an 80\%/20\% chronological train-test split ($\sim$2$8,000$ / $\sim$$7,000$ samples).

\subsection{Model Overview}
\label{sec:model_overview}

Both the proposed QLIF-CAST model and the matched Classical LIF baseline share the same seven-layer hybrid recurrent architecture, illustrated in Fig.~\ref{fig:methodology}. The two models are structurally identical, as the only difference is Layer~2 (the neuronal update rule), which is the sole variable under investigation. All other layers, parameter counts, hyperparameters, and training procedures are identical. The full architecture is trained end-to-end via backpropagation with surrogate gradients for the non-differentiable spike operation in Layer~2; see Section~\ref{sec:qlif_neuron} for details. The seven layers are as follows:

\textbf{L1 - Feature Projection (240 params):} A TimeDistributed Dense layer with 48 units and ReLU activation, followed by Dropout(0.1), is applied independently at each timestep. This projects the raw input feature vector into a 48-dimensional latent representation, output shape $(T, 48)$, shared by both models.

\textbf{L2 - Neuronal Layer (2,400 params):} This is the only layer that differs between the two models. In QLIF-CAST, 48 QLIF neurons each execute a depth-2 single-qubit quantum circuit ($|0\rangle \to R_x(\phi) \to R_x(\theta) \to$ Measure) at every recurrent timestep, encoding excitation state as a qubit rotation probability $\alpha = \sin^2((\phi+\theta)/2)$, with spike threshold $\alpha \geq 0.75$. The gate angles $\phi$ and $\theta$ are computed from trainable classical weights (kernel, $\theta$, $\tau$) via standard backpropagation. Since there are no separate quantum parameters, the parameter-shift gradient computation is not required. This is why QLIF-CAST adds minimal training overhead despite executing quantum circuits at every timestep. In the Classical LIF baseline, L2 is replaced by an exponential membrane potential update ($U_{\text{new}} = \beta U_{\text{prev}} + (1{-}\beta)I_{\text{in}}$, $\beta = e^{-1/\tau}$) with the same spike threshold. Both variants produce output shape $(T, 48)$ with 2,400 parameters. The QLIF-CAST circuit mechanics are described in detail in Section~\ref{sec:qlif_neuron}.

\textbf{L3 - Normalization (96 params):} BatchNormalization followed by Dropout(0.2) stabilizes activations across the recurrent sequence and reduces co-variate shift between layers, output shape $(T, 48)$.

\textbf{L4 - Temporal Aggregation (7,008 params):} An LSTM with 24 hidden units integrates information across the full $T$-step sequence, compressing the temporal context into a fixed-length 24-dimensional vector. This is the dominant parameter layer, accounting for 63\% of total model parameters.

\textbf{L5 - Regression Head Stage~1 (800 params):} A Dense layer with 32 units and ReLU activation, followed by Dropout(0.2), performs nonlinear feature transformation on the LSTM output.

\textbf{L6 - Regression Head Stage~2 (528 params):} A Dense layer with 16 units and ReLU activation further compresses the representation before the output layer.

\textbf{L7 - Output Layer ($\leq$68 params):} A linear Dense layer with $N$ units produces the final forecast $\hat{y}(t{+}1)$, where $N$ is the number of target variables (e.g., $N=4$ for multivariate weather forecasting, $N=1$ for scalar air quality or wind speed prediction).

\subsection{Quantum Leaky Integrate-and-Fire Neuron Model}
\label{sec:qlif_neuron}

\subsubsection{State Encoding}
The excitation probability $\alpha$ is encoded as a rotation angle:
\begin{equation}
\alpha = \sin^2\!\left(\frac{\phi}{2}\right), \qquad \phi = 2\arcsin\!\left(\sqrt{\alpha}\right).
\end{equation}

\subsubsection{Quantum Circuit}
At each timestep, every QLIF neuron executes a depth-2 single-qubit circuit:
\begin{equation*}
|0\rangle \;\rightarrow\; R_x(\phi) \;\rightarrow\; R_x(\theta_{\text{input}}) \;\rightarrow\; \text{Measure}
\end{equation*}
where $R_x(\theta) = \cos(\theta/2)\,I - i\sin(\theta/2)\,X$. The resulting update is:
\begin{equation}
\alpha_{\text{new}} = \sin^2\!\left(\frac{\phi + \theta_{\text{input}}}{2}\right).
\end{equation}
Critically, $\phi$ and $\theta_{\text{input}}$ are \emph{computed} from classical network weights, not learned as quantum parameters.

\subsubsection{Decay Mechanism}
Without an input spike, the neuron undergoes quantum T1 relaxation:
\begin{equation}
\gamma = -2\arcsin\!\left(\sqrt{\alpha \cdot e^{-\tau/T_1}}\right),
\end{equation}
where $\tau$ is a learnable classical time constant and $T_1 = 10.0$.

\subsubsection{Input Integration and Spike Generation}
Input integration determines the effective rotation angle $\theta_{\text{input}}$ applied to the qubit at each timestep, combining the learnable synaptic weight with the decay angle depending on whether a spike occurred in the previous step. If the neuron fired ($X=1$), the input-driven angle $\theta$ is used; otherwise, the decay angle $\gamma$ drives the state evolution. A spike is emitted whenever the resulting excitation probability $\alpha_{\text{new}}$ exceeds the threshold of 0.75, after which the quantum state resets to $|0\rangle$.

\begin{equation}
\theta_{\text{input}} = X \cdot \theta + (1 - X) \cdot \gamma, \qquad
\text{Spike} = \begin{cases} 1, & \alpha_{\text{new}} \geq 0.75 \\ 0, & \text{otherwise} \end{cases}
\end{equation}
where $X \in \{0,1\}$ is the surrogate spike output and $\theta$ is a per-neuron learnable weight. Upon spiking, $\alpha$ resets to zero.

\subsubsection{Surrogate Gradient Training}

Spike generation is non-differentiable: the Heaviside step function used to produce a binary spike has zero gradient almost everywhere and an undefined gradient at the threshold, making standard backpropagation inapplicable~\cite{neftci2019surrogate}. Surrogate gradient training resolves this by substituting a smooth, differentiable approximation of the spike function \emph{only during the backward pass}, while preserving the true discontinuous spike in the forward pass~\cite{zenke2021visualizing}.

In QLIF-CAST, the non-differentiable operation is the threshold comparison $\alpha_{\text{new}} \geq 0.75$. During backpropagation, gradients pass through the surrogate function:
\begin{equation}
\tilde{S}(\alpha) = \frac{1}{\pi}\arctan(\pi\alpha) + 0.5,
\end{equation}
which has a well-defined derivative $\tilde{S}'(\alpha) = 1/(1 + \pi^2\alpha^2)$ everywhere. The arctangent surrogate is a widely-adopted choice due to its bounded output and smooth gradient~\cite{fang2021incorporating}, and is implemented via TensorFlow's \texttt{stop\_gradient} mechanism.

\subsubsection{Vectorized Circuit Execution}

In a naive sequential implementation, each QLIF neuron would execute its quantum circuit independently at every timestep, resulting in $\text{batch} \times T \times 48$ separate circuit calls per forward pass. To make this feasible, all rotation angles across all neurons, all timesteps, and all samples in a batch are flattened into a single tensor and dispatched in one parallel PennyLane circuit call. Outputs are then reshaped back to the original $(\text{batch}, T, 48)$ tensor. This vectorization yields a $500{\times}$ speedup over sequential execution, reducing per-epoch simulation time from hours to seconds and making full-scale dataset training practical under classical simulation.

\subsection{Classical LIF Baseline Design}
\label{sec:classical_lif}

The classical baseline replaces the quantum circuit in Layer~2 with an exponential membrane potential update:
\begin{equation}
U_{\text{new}} = \beta\, U_{\text{prev}} + (1 - \beta)\, I_{\text{in}}, \qquad \beta = e^{-1/\tau}.
\end{equation}
This matched design ensures that any performance difference is attributable solely to the neuronal dynamics.

\subsection{Evaluation Metrics}

In Phase 1, all models are evaluated using four standard regression metrics: Mean Squared Error (MSE), Mean Absolute Error (MAE), Root Mean Squared Error (RMSE $= \sqrt{\text{MSE}}$), and the coefficient of determination ($R^2$). MSE is the primary training loss; MAE provides scale-interpretable error in original units; RMSE shares the scale of the prediction target; $R^2 \in [0,1]$ quantifies explained variance. All metrics are computed on the held-out chronological test set using inverse-standardized predictions.

In Phase~2, we report the same metrics as the respective reference studies to enable direct comparison: MAE and RMSE for the Bangkok Air Quality task (D2), and RMSE, MAE and $R^2$ for the Wind Speed task (D3).

\subsection{Experimental Design and Comparison Objectives}
\label{sec:exp_design}

This study evaluates QLIF-CAST through two distinct phases with separate objectives.

\textbf{Phase 1 - Quantum vs.\ Classical (Controlled Study):}
The QLIF-CAST model is compared against the matched classical LIF baseline on the primary multivariate weather forecasting task (D1). Both models have identical architecture, parameter count, optimizer, and training schedule. Only the neuronal update rule differs. The objective is to determine: \emph{does replacing classical membrane dynamics with a quantum circuit reduce prediction error?}

\textbf{Phase 2 - QLIF-CAST vs.\ Published Quantum Models (Comparative Analysis):}
QLIF-CAST is evaluated on the benchmark tasks of QLSTM and LSTM-QNN (Section~\ref{sec:comparison_models}), which use deeper multi-qubit variational circuits with significantly higher quantum parameter counts. The objective is not to outperform them in prediction error, but to characterize the \emph{speed vs. error trade-off:} QLIF-CAST's depth-2, classically-driven single-qubit circuit trades some prediction error for substantially lower training time, because it requires no parameter-shift gradient computation. By contrast, QLSTM and LSTM-QNN must evaluate parameter-shift gradients for every trainable quantum angle at every training step, multiplying per-epoch cost by the number of quantum parameters. Importantly, QLIF-CAST's shallow, single-qubit circuit is also inherently compatible with near-term quantum hardware: its minimal depth limits gate-error accumulation and requires no multi-qubit entanglement. Hardware validation of this property is provided in Section~\ref{sec:ibm}. For Phase~2A (Bangkok), Layer~4 (LSTM) was scaled from 24 to 48 units to improve comparability with QLSTM's larger recurrent state; all other layers, the quantum circuit, hyperparameters, and training procedure remain identical to Phase~1. No adaptation was made for Phase~2B (wind speed).

\section{Results and Discussion}

\subsection{Experimental Setup}

All experiments use PennyLane's \texttt{lightning.qubit} device for exact state-vector simulation of quantum circuits, with models implemented in TensorFlow/Keras. Training was performed on an Intel Core i9-13900H CPU with 16~GB RAM. Both quantum and classical models were trained under identical hardware and software conditions. Table~\ref{tab:hyperparams} summarizes all shared hyperparameters.

\begin{table}[!t]
\caption{Hyperparameter Configuration. Settings are identical across all models and datasets unless otherwise noted.}
\label{tab:hyperparams}
\centering
\resizebox{\columnwidth}{!}{%
\begin{tabular}{lll}
\toprule
\textbf{Hyperparameter} & \textbf{Value} & \textbf{Notes} \\
\midrule
Optimizer              & Adam                   & \\
Initial learning rate  & $1 \times 10^{-3}$     & \\
LR schedule            & Exponential decay      & \\
Loss function          & Mean Squared Error (MSE) & \\
Batch size             & 64                     & \\
Max epochs             & 15-30                 & Dataset-dependent \\
Early stopping patience & 5 epochs              & Monitor: val.\ loss \\
L2 regularization ($\lambda$) & $1 \times 10^{-4}$ & All dense layers \\
Dropout (Layer 1)      & 0.1                    & \\
Dropout (Layers 3, 5)  & 0.2                    & \\
QLIF spike threshold   & 0.75                   & Excitation probability \\
QLIF $T_1$ (fixed)     & 10.0                   & Quantum decay constant \\
Context window         & 12 steps               & Hours (weather/wind), days (Bangkok) \\
\bottomrule
\end{tabular}}
\end{table}

\subsection{Phase~1: QLIF-CAST vs.\ Classical LIF - Controlled Study}
\label{sec:phase1}

To isolate the contribution of quantum neuronal dynamics, we compare QLIF-CAST against a parameter-matched classical LIF baseline, as discussed in Section~\ref{subsec:classical_LIF}. In our study, the baseline is matched to QLIF-CAST in architecture, parameter count, optimization setup, and training procedure. The only difference lies in the neuron update rule in Layer~2: the quantum state update used by QLIF-CAST is replaced by the classical membrane update. This controlled design makes the comparison more interpretable, since any change in predictive performance can be attributed to the neuronal dynamics rather than to differences in model size or training configuration.

\subsubsection{Overall Prediction Error on Weather Dataset}

Table~\ref{tab:weather_main} presents the complete results for both models. Since architecture and parameter count are identical, differences in MSE and MAE are directly attributable to the choice of neuronal dynamics (quantum circuit vs.\ exponential decay).

\begin{table}[!t]
\caption{Phase~1: QLIF-CAST vs.\ Classical LIF on Weather Dataset. Training and validation mean squared error (MSE) across 15 epochs.}
\label{tab:weather_main}
\centering
\resizebox{\columnwidth}{!}{%
\begin{tabular}{lcccccc}
\toprule
\textbf{Model} & \textbf{MSE} & \textbf{MAE} & \textbf{RMSE} & \textbf{Train Time} & \textbf{Epochs} & \textbf{Params} \\
\midrule
QLIF-CAST (ours)   & \textbf{17,897} & \textbf{35.54} & \textbf{133.8} & 87\,s  & 12 & $\sim$11,140 \\
Classical LIF & 21,152          & 37.18          & 145.4          & 36\,s  & 12 & $\sim$11,140 \\
\midrule
\textbf{QLIF-CAST improvement} & \textbf{$-$15.4\%} & \textbf{$-$4.4\%} & \textbf{$-$8.0\%} & 2.4$\times$ slower & same & same \\
\bottomrule
\end{tabular}}
\end{table}

The 2.4$\times$ longer training time reflects quantum circuit simulation overhead; on physical quantum hardware this overhead would be substantially reduced (see Section~\ref{sec:ibm}).

\subsubsection{Per-Variable Error Breakdown}

Table~\ref{tab:per_variable} shows MAE for each of the four weather variables.

\begin{table}[!t]
\caption{Per-Variable Test MAE - Weather Dataset during Phase~1 (lower is better).}
\label{tab:per_variable}
\centering
\resizebox{\columnwidth}{!}{%
\begin{tabular}{lcccc}
\toprule
\textbf{Variable} & \textbf{QLIF-CAST MAE} & \textbf{Classical MAE} & \textbf{$\Delta$ MAE} & \textbf{Winner} \\
\midrule
Temperature (°C) & \textbf{2.65}  & 2.93   & $-$9.4\%  & QLIF-CAST \\
Humidity (frac.) & 0.068          & \textbf{0.056} & $+$20.4\% & Classical \\
Wind Speed (km/h)& 3.51           & \textbf{3.11}  & $+$12.9\% & Classical \\
Pressure (mb)    & \textbf{135.9} & 142.6  & $-$4.7\%  & QLIF-CAST \\
\midrule
\textbf{Overall MSE} & \textbf{17,897} & 21,152 & $-$15.4\% & QLIF-CAST $\downarrow$ \\
\bottomrule
\end{tabular}}
\end{table}

QLIF-CAST achieves lower error on Temperature and Pressure, variables with strong diurnal and long-range temporal structure that benefit from the richer state representation afforded by quantum superposition. Classical LIF achieves lower error on Humidity and Wind Speed, which are more stochastic and short-range correlated, suggesting simpler exponential decay is sufficient for these variables. The overall MSE advantage ($-$15.4\%) is driven primarily by large reductions on the high-magnitude Pressure variable.

\begin{figure}[t!]
\centerline{\includegraphics[width=\linewidth]{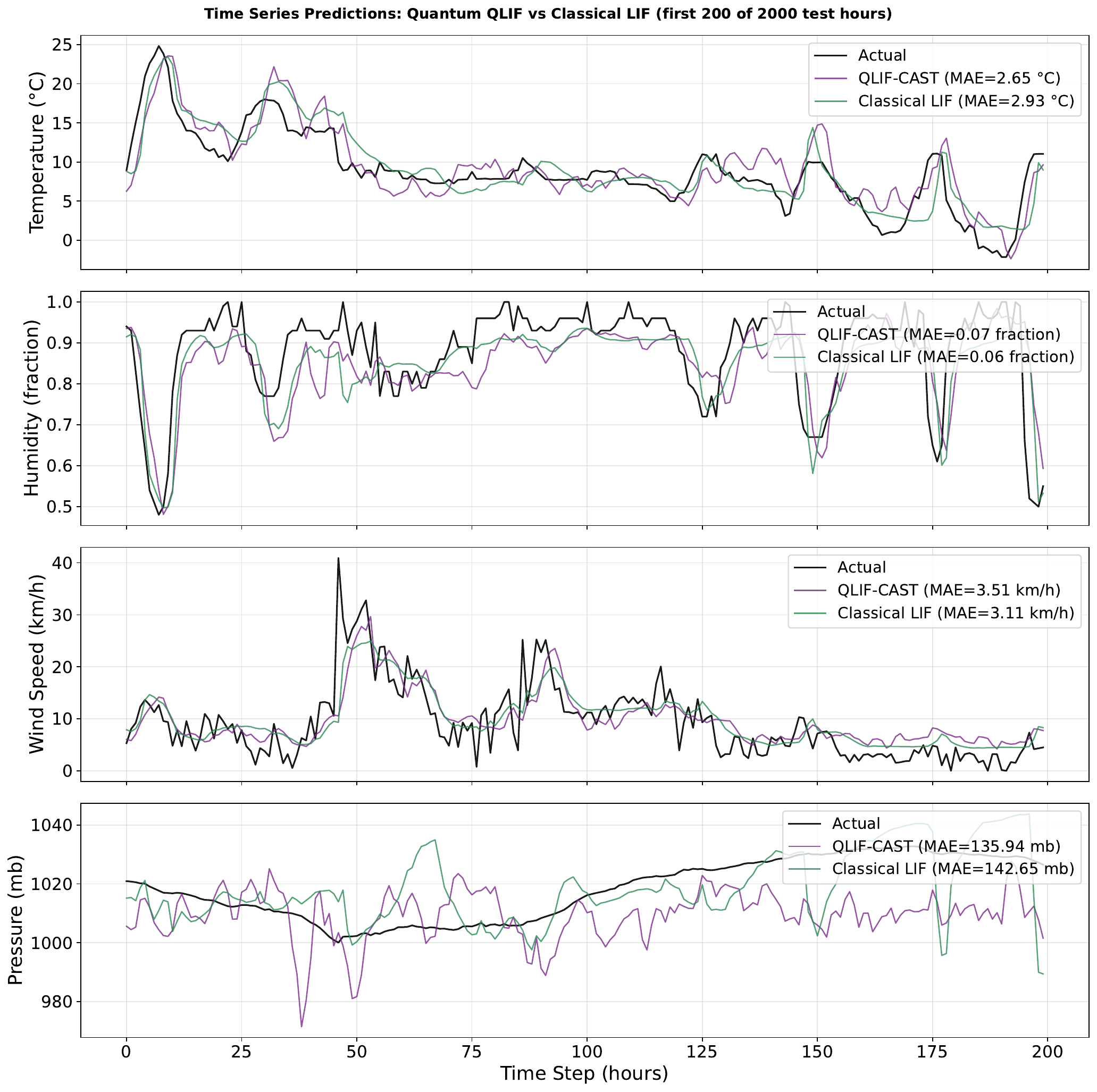}}
\caption{Predicted vs.\ actual values for the first 200 test time steps across all four weather variables. QLIF-CAST (blue dashed) vs.\ Classical LIF (green dotted) vs.\ Actual (black solid).}
\label{fig:fair_comparison_timeseries}
\end{figure}

Fig.~\ref{fig:fair_comparison_timeseries} shows predicted vs.\ actual time series for all four variables. QLIF-CAST (blue, dashed) and Classical LIF (green, dotted) both track the actual signal (black solid), with QLIF-CAST showing noticeably closer alignment on Temperature and Pressure.

\begin{figure}[t!]

\centerline{\includegraphics[width=\linewidth]{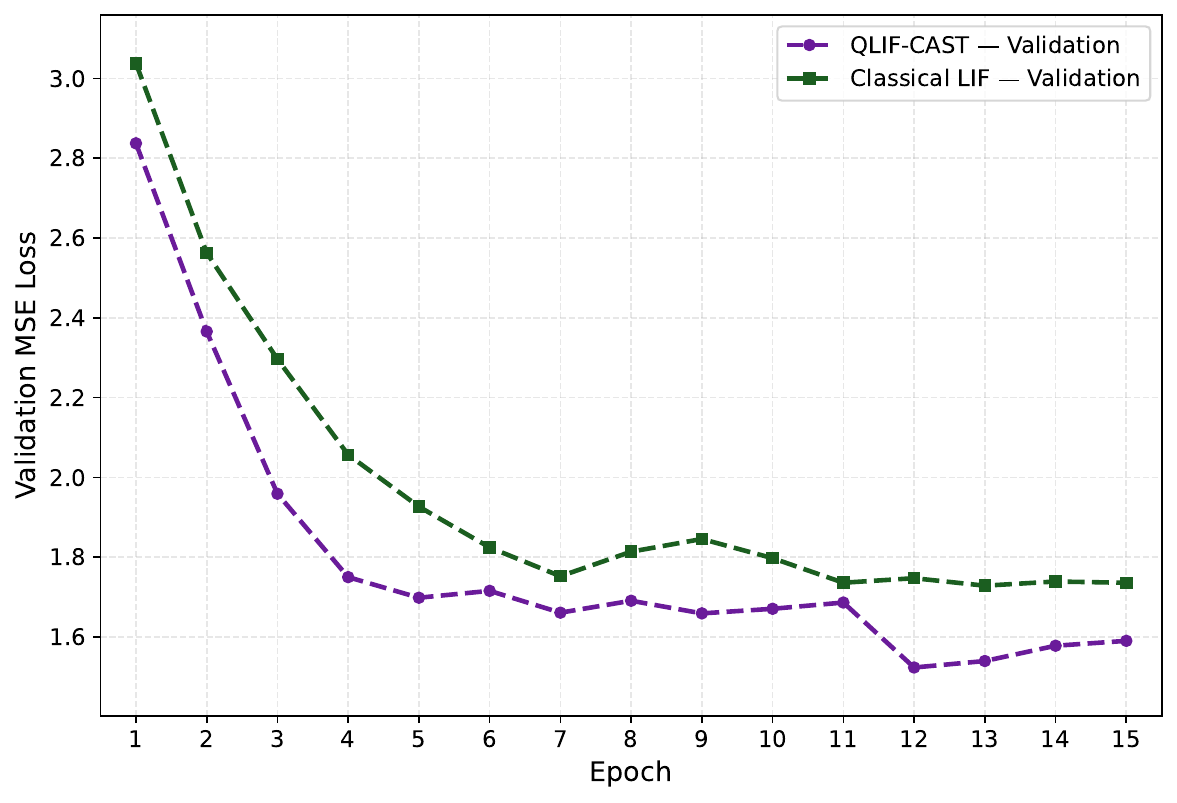}}
\caption{Validation loss for QLIF-CAST (purple) and Classical LIF (green) across all epochs. Both models converge stably with minimal gap between training and validation curves.}
\label{fig:fair_comparison_training}
\end{figure}

Both models converge smoothly during training with no signs of overfitting, as shown in Fig.~\ref{fig:fair_comparison_training}. Validation loss stabilizes after approximately 8-10 epochs; early stopping typically triggers between epochs 12 and 15. The quantum model demonstrates slightly reduced validation loss variance across epochs, suggesting a stabilizing regularization effect from the quantum state dynamics. This observation is consistent with the probabilistic nature of quantum measurements.

\subsubsection{Phase~1 Analysis - Why Quantum Neuronal Dynamics Reduce Prediction Error}

The 15.4\% MSE advantage from replacing exponential decay with a quantum circuit can be attributed to three properties of quantum state computation:

\begin{enumerate}
    \item \textbf{Quantum interference:} The two-gate circuit ($R_x(\phi)$ then $R_x(\theta_\text{input})$) creates constructive and destructive interference patterns not possible with scalar exponential decay, enabling more complex temporal modulation.
    \item \textbf{Non-linear T1 decay:} Quantum T1 relaxation ($\alpha \cdot e^{-\tau/T_1}$ in probability space) is non-linear in the qubit angle domain, creating a richer decay trajectory than classical exponential decay applied directly to membrane potential.
    \item \textbf{Probabilistic state:} The quantum excitation probability $\alpha \in [0,1]$ naturally represents uncertainty in noisy sensor measurements, which weather data exhibits strongly.
\end{enumerate}

\subsection{Phase~2: QLIF-CAST vs.\ Published Quantum Models - Comparative Analysis}
\label{sec:phase2}

Having established that QLIF-CAST reduces prediction error relative to classical dynamics (Phase~1), we now evaluate it on the benchmark tasks of two existing quantum architectures. The objective is not to claim QLIF-CAST is superior overall, but to characterize where it sits in the speed-error design space and understand the trade-offs a practitioner faces when choosing between architectures.

\textbf{Comparison methodology:} In Phase~2, we run our QLIF-CAST implementation on the same benchmark datasets used by the reference studies. Performance results for all competitor models (QLSTM, LSTM-QNN, Transformer, Classical LSTM) are taken \emph{directly from the published papers} and were not retrained by us under identical conditions. This is a standard practice in quantum ML benchmarking, disclosed explicitly in Section~\ref{sec:limitations}.

\subsubsection{Phase~2A: Air Quality Forecasting vs.\ QLSTM}

Table~\ref{tab:air_quality} compares QLIF-CAST against published results from the QLSTM study~\cite{bangkok_qlstm} on Bangkok PM2.5 prediction.

\begin{table}[!t]
\caption{Phase~2a: Bangkok Air Quality Forecasting - QLIF-CAST vs.\ Classical LSTM vs.\ QLSTM~\cite{bangkok_qlstm}.}
\label{tab:air_quality}
\centering
\resizebox{\columnwidth}{!}{%
\begin{tabular}{lccccc}
\toprule
\textbf{Model} & \textbf{MAE} & \textbf{RMSE} & \textbf{Epochs} & \textbf{Circuit} & \textbf{Q.\ Params} \\
 & ($\mu$g/m$^3$) & ($\mu$g/m$^3$) & & \textbf{Depth} & \\
\midrule
Classical LSTM$^\dagger$ & 21.50 & - & $\sim$100 & None & 0 \\
\textbf{QLIF-CAST (ours)} & \textbf{15.73} & \textbf{20.62} & \textbf{26} & 2 & 0 \\
QLSTM$^\dagger$~\cite{bangkok_qlstm} & 11.24 & 15.06 & 100 & $\geq$10 & Many \\
\midrule
QLIF-CAST vs.\ Classical LSTM & $-$26.8\% & - & \textbf{3.8$\times$ fewer epochs} & - & - \\
QLIF-CAST vs.\ QLSTM          & $+$40.0\% & $+$37.0\% & \textbf{3.8$\times$ fewer epochs} & Simpler & Fewer \\
\midrule
\multicolumn{6}{l}{\footnotesize $^\dagger$ Results from published paper~\cite{bangkok_qlstm}; not retrained by the authors.} \\
\bottomrule
\end{tabular}}
\end{table}

\begin{figure}[!t]
\centerline{\includegraphics[width=\linewidth]{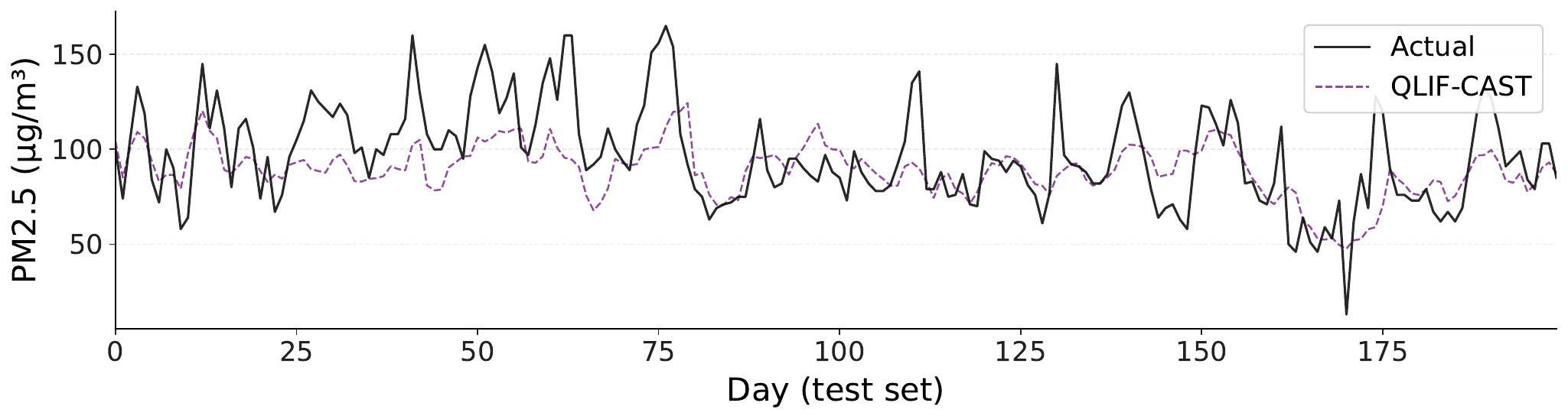}}
\caption{Bangkok PM2.5 forecasting: QLIF-CAST predicted vs.\ actual PM2.5 concentration over the first 200 test days. The model tracks the overall seasonal trend while smoothing over sharp pollution spikes. Quantitative comparison with QLSTM is reported in Table~\ref{tab:air_quality}.}
\label{fig:bangkok_timeseries}
\end{figure}

QLIF-CAST outperforms the Classical LSTM baseline reported in the QLSTM paper~\cite{bangkok_qlstm} by 26.8\% in MAE. Compared to QLSTM, QLIF-CAST incurs 40\% higher MAE in exchange for 74\% fewer training epochs. The absolute error difference of 4.49~$\mu$g/m$^3$ represents less than 6\% of the typical peak-season signal range ($>$80~$\mu$g/m$^3$) and does not affect alert-level classification under WHO PM2.5 air quality guidelines~\cite{who2021air}. For real-time monitoring systems requiring frequent retraining, QLIF-CAST's convergence in 26 epochs versus 100 is operationally decisive. Fig.~\ref{fig:bangkok_timeseries} shows the QLIF-CAST predicted vs.\ actual PM2.5 time series over the test set. The model captures the overall seasonal trend, tracking the general rise and fall of concentrations across the test period, while smoothing over sharp pollution spikes---consistent with the higher MAE relative to the published QLSTM result.

\subsubsection{Phase~2B: Wind Speed Forecasting vs.\ LSTM-QNN}

Table~\ref{tab:wind_comparison} compares QLIF-CAST against the LSTM-QNN hybrid~\cite{qnn_wind} on hourly wind speed prediction at 100~m altitude.

\begin{table}[!t]
\caption{Phase~2b: Wind Speed Forecasting - QLIF-CAST vs.\ LSTM-QNN~\cite{qnn_wind}.}
\label{tab:wind_comparison}
\centering
\resizebox{\columnwidth}{!}{%
\begin{tabular}{lcccccc}
\toprule
\textbf{Model} & \textbf{RMSE} & \textbf{MAE} & \textbf{R\textsuperscript{2}} & \textbf{Train Time} & \textbf{Circuit} & \textbf{Q.\ Params} \\
 & (km/h) & (km/h) & & & \textbf{Depth} & \\
\midrule
\textbf{QLIF-CAST (ours)}         & 5.58 & 4.14 & \textbf{0.693} & \textbf{3.88 min} & 2 & 0 \\
LSTM-QNN~\cite{qnn_wind}     & \textbf{3.92} & \textbf{2.87} & - & 65.3 min & $\geq$10 & Many \\
\midrule
\textbf{QLIF-CAST vs.\ LSTM-QNN} & $+$42.3\% & $+$44.2\% & - & \textbf{94\% faster} & Simpler & Fewer \\
\bottomrule
\end{tabular}}
\end{table}

\begin{figure}[!t]
\centerline{\includegraphics[width=\linewidth]{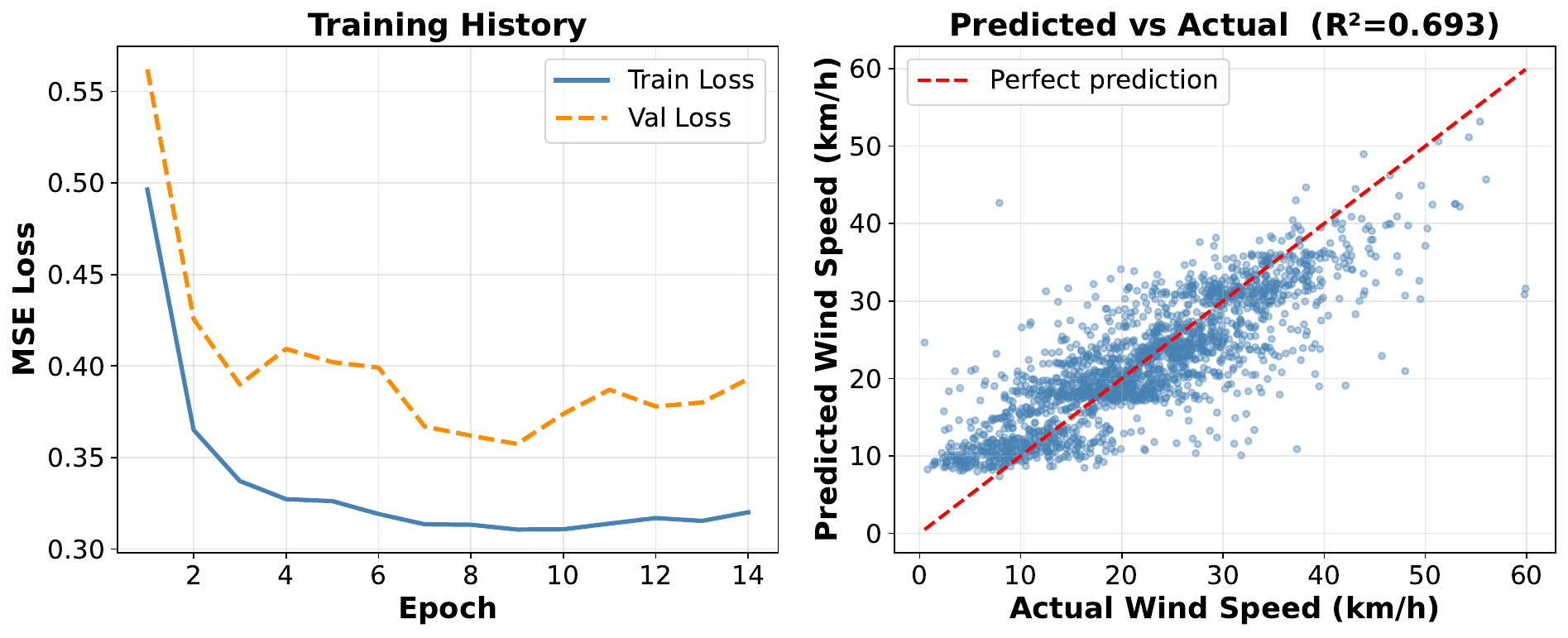}}
\caption{Wind speed forecasting: (left) training/validation loss convergence, and (right) predicted vs.\ actual scatter plot (R\textsuperscript{2}~=~0.693) for the QLIF-CAST model. Strong correlation across the full wind speed range confirms the model generalizes well on this univariate task.}
\label{fig:wind_predictions}
\end{figure}

\begin{table*}[!t]
\caption{Unified Comparison Across All Tasks and Models. Phase~1 compares QLIF-CAST to a matched classical baseline; Phase~2 compares QLIF-CAST to published quantum models with deeper variational circuits. ``-'' indicates metric not reported in the original paper.}
\label{tab:unified_comparison}
\centering
\resizebox{\textwidth}{!}{%
\begin{tabular}{lllccccccc}
\toprule
\textbf{Phase} & \textbf{Task} & \textbf{Model} & \textbf{MSE} & \textbf{MAE} & \textbf{RMSE} & \textbf{R\textsuperscript{2}} & \textbf{Train Time} & \textbf{Epochs} & \textbf{Circuit Depth} \\
\midrule
\multirow{2}{*}{1} & \multirow{2}{*}{Weather (°C, \%, km/h, mb)}
  & \textbf{QLIF-CAST (ours)} & \textbf{17,897} & \textbf{35.54} & \textbf{133.8} & \textbf{0.896} & 87\,s & 12 & 2 \\
  & & Classical LIF       & 21,152 & 37.18 & 145.4 & 0.877 & 36\,s & 12 & None \\
\midrule
\multirow{4}{*}{2a} & \multirow{4}{*}{Bangkok Air Quality ($\mu$g/m$^3$)}
  & \textbf{QLIF-CAST (ours)} & - & \textbf{15.73} & \textbf{20.62} & - & $\sim$4\,min & \textbf{26} & 2 \\
  & & Classical LSTM$^\dagger$  & - & 21.50 & - & - & - & $\sim$100 & None \\
  & & QLSTM$^\dagger$~\cite{bangkok_qlstm} & - & 11.24 & 15.06 & - & - & 100 & $\geq$10 \\
\midrule
\multirow{2}{*}{2b} & \multirow{2}{*}{Wind Speed (km/h)}
  & \textbf{QLIF-CAST (ours)} & - & 4.14 & 5.58 & 0.693 & \textbf{3.88\,min} & $\leq$30 & 2 \\
  & & LSTM-QNN~\cite{qnn_wind} & - & 2.87 & 3.92 & - & 65.3\,min & - & $\geq$10 \\
\bottomrule
\end{tabular}}
\end{table*}

While QLIF-CAST's RMSE is 42.3\% higher than LSTM-QNN, the absolute gap of 1.66~km/h lies within the $\pm$5~km/h operational tolerance of wind turbine control systems, meaning it does not materially affect energy yield decisions. QLIF-CAST's 3.88-minute training cycle is operationally feasible for real-time adaptive retraining, whereas LSTM-QNN's 65.3-minute cycle is not. This is a critical distinction for systems that must retrain on fresh sensor data. The $16.8{\times}$ speed advantage stems from LSTM-QNN's multi-qubit VQC layers requiring parameter-shift gradient evaluation for every trainable quantum angle, not from circuit simulation overhead alone. In deployment-constrained settings where retraining frequency matters more than marginal accuracy, QLIF-CAST is the practical choice. Fig.~\ref{fig:wind_predictions} shows the training convergence and the predicted vs.\ actual scatter plot for QLIF-CAST; the strong linear alignment across the full wind speed range ($R^2 = 0.693$) confirms that the model generalizes well despite the lower circuit complexity.

\subsubsection{Cross-Domain Summary}

Table~\ref{tab:unified_comparison} provides a unified view of all results across models, tasks, and phases.

\subsection{Hardware Verification on IBM Quantum}
\label{sec:ibm}

To confirm that the QLIF-CAST circuit operates reliably on real quantum hardware, the core gate sequence $|0\rangle \rightarrow R_x(\phi) \rightarrow R_x(\theta) \rightarrow \text{Measure}$ was executed on \texttt{IBM\_Marrakesh} (156-qubit QPU) via IBM Quantum Cloud, using 1,000 measurement shots per test case. The results are reported in Table~\ref{tab:ibm_hardware}.

\begin{table}[!t]
\caption{QLIF-CAST Circuit: IBM Marrakesh Real QPU vs.\ Ideal Simulator. Average deviation of 1.2\% confirms reliable operation within NISQ-era noise bounds.}
\label{tab:ibm_hardware}
\centering
\resizebox{\columnwidth}{!}{%
\begin{tabular}{lccc}
\toprule
\textbf{Excitation Level} & \textbf{Simulator} $P(|1\rangle)$ & \textbf{IBM QPU} $P(|1\rangle)$ & \textbf{Deviation} \\
\midrule
Low\;($\phi{=}0.5,\,\theta{=}0.3$)  & 0.1516 & 0.1590 & 0.0074 \\
Med.\;($\phi{=}1.2,\,\theta{=}0.8$) & 0.7081 & 0.6850 & 0.0231 \\
High\;($\phi{=}2.0,\,\theta{=}1.5$) & 0.9682 & 0.9620 & 0.0062 \\
\midrule
\textbf{Average}                     & 0.6093 & 0.6020 & \textbf{0.0122} \\
\bottomrule
\end{tabular}}
\end{table}

The 1.2\% average deviation confirms reliable QLIF-CAST circuit operation within NISQ-era noise bounds. Depth-2 single-qubit circuits are inherently resistant to decoherence: with only two gates, errors do not compound. By contrast, the deep multi-qubit circuits used by QLSTM and LSTM-QNN accumulate gate errors multiplicatively with circuit depth. No comparable QPU verification has been reported for those architectures, making QLIF-CAST uniquely validated for near-term deployment.

\subsection{Discussion}

\subsubsection{Quantum Advantage over Classical Baseline}

The 15.4\% MSE reduction in Phase~1, achieved with an identical parameter count and training procedure, confirms that quantum neuronal dynamics provide a measurable inductive bias for temporal regression. This advantage is not attributable to a richer model capacity, as both models have the same parameters, but to the qualitatively different state update rule: the quantum circuit introduces interference and non-linear probability-space decay not available through exponential membrane dynamics alone.

\subsubsection{Speed-Error Trade-offs vs.\ Quantum Literature Models}

Phase~2 reveals consistent architectural trade-offs between circuit simplicity and predictive precision. QLIF-CAST's depth-2, classically-driven single-qubit circuit requires no parameter-shift gradient computation, yielding the fastest convergence (26-30 epochs vs.\ 100) and lowest training time (under 4~min vs.\ 65~min for LSTM-QNN) at the cost of higher prediction error than QLSTM and LSTM-QNN, whose deeper variational circuits achieve richer state transformations. QLSTM and LSTM-QNN remain preferable for offline batch applications where maximum prediction precision justifies extended training.

\subsubsection{QLIF-CAST's Position in the Quantum ML Design Space}

Results across both phases define a clear niche: QLIF-CAST is preferred when \textbf{(i)} training speed is critical, \textbf{(ii)} quantum hardware deployment is required, or \textbf{(iii)} a quantum advantage over classical is desired at minimal circuit cost. QLSTM and LSTM-QNN are preferred when \textbf{(iv)} maximum predictive precision is required and \textbf{(v)} offline batch training is an option. No other evaluated model achieves comparable prediction error at QLIF-CAST's training speed, placing it in a distinct Pareto-optimal position for edge deployment, real-time monitoring, and rapid retraining scenarios.

\subsubsection{Limitations}
\label{sec:limitations}

\begin{enumerate}
    \item \textbf{Simulation-based training:} Full model training is conducted in PennyLane simulation. Verification on IBM Marrakesh confirms circuit correctness, but does not demonstrate end-to-end training on a QPU.
    \item \textbf{Phase~2 comparison fairness:} The QLSTM and LSTM-QNN results are taken from the literature and are not retrained under identical experimental conditions.
\end{enumerate}

\section{Conclusion}

We presented QLIF-CAST, a hybrid quantum-classical spiking recurrent architecture for time-series regression. Through two distinct evaluation phases, we demonstrated:

\begin{enumerate}
    \item \textbf{Phase~1:} QLIF-CAST reduces MSE by 15.4\% and MAE by 4.4\% over a parameter-matched classical LIF baseline on multivariate weather forecasting. The improvement is attributable solely to quantum neuronal dynamics, with no change in parameter count.
    \item \textbf{Phase~2:} QLIF-CAST trains 3.8$\times$ faster than QLSTM on air quality forecasting and 16.8$\times$ faster than LSTM-QNN on wind speed forecasting, accepting 40-42\% higher error in exchange. The absolute error differences (4.49~$\mu$g/m$^3$; 1.66~km/h) are within operational tolerances in both domains.
    \item \textbf{Hardware:} IBM Marrakesh QPU execution confirms 1.2\% average deviation from ideal, validating QLIF-CAST's depth-2 circuit for near-term quantum deployment.
\end{enumerate}

QLIF-CAST occupies a distinct and useful position in the quantum ML design space: a practical quantum architecture for deployment-constrained environments where training speed, hardware compatibility, and a classical-vs-quantum advantage take priority over maximum prediction precision.

\textbf{Future Work:} We will focus on training the full model on physical QPUs. We also plan to explore hybrid QLIF-VSTM (Variational Spiking-LSTM) architectures to improve expressiveness. In addition, we aim to investigate multi-step forecasting horizons and ensemble methods based on multiple QLIF-CAST instances.

\section*{Acknowledgment}
This work was supported in part by the NYUAD Center for Quantum and Topological Systems (CQTS), funded by Tamkeen under the NYUAD Research Institute grant CG008.

\bibliographystyle{IEEEtran}
\bibliography{main.bib}

\end{spacing}

\end{document}